\documentclass[prl,twocolumn,superscriptaddress]{revtex4}
\usepackage{graphicx}
\usepackage{graphics}      
\usepackage{bm}    
\usepackage{amsmath}     
\usepackage{amsfonts}
\usepackage{amssymb}
\usepackage{latexsym}  
\begin{document}
\title{A Pedagogical Approach to Quantum Computing using spin-1/2 particles}
\newcounter{count}
\author{Prashant}
\email{prashant.iiitm@gmail.com} \affiliation{Indian Institute of IT
and Management, India.}
\author{I.Chakrabarty}
\email{indranilc@indiainfo.com} \affiliation{Heritage Institute of
Technology, Kolkata, India} \affiliation{Bengal Engineering and
Science University, Howrah, India}
\date{\today}
\begin{abstract}
\textbf{Abstract:} This paper discusses the important primitives of
superposition and entanglement in QIP from physics of spin-1/2
particles. System of spin-1/2 particles present a logical and
conceptual candidate to understand Quantum Computing. A pedagogical
approach to abstract quantum information processing is considered in
more concrete physical terms here.
\end{abstract}
\maketitle
\section{1. Introduction}
Quantum  information  processing  (QIP) is set to change computation
and   communication   in   a  way  inconceivable  by  our  classical
perceptions. QIP derives its power, not from any revolutionary
advances  in  the existing technologies, but from the logical system
based on the quantum mechanical formalism. The quantum theoretic
concepts on which  QIP  stands  are  superposition  and
entanglement.
\paragraph{}
Superposition  refers  to  the possibility of a system existing as a
linear  combination  of different states commensurate with the given
conditions.  It  manifests  in  the  form of observable interference
effects  between  those  states  even  in  the particle description.
Recall that in classical mechanics, interference is a characteristic
of        waves        not        expected       of       particles.
\paragraph{}
Entanglement  refers  to correlated states of two or more particles.
There is nothing quantum mechanical about correlations per se for
example,  consider  an  experiment  in  which pairs of particles are
created such that their total linear momentum is zero. Hence, if the
momentum  of one of the particles of a pair is found to be \emph{p}
in some direction then that of the other in that direction will
certainly be found  to  be  \emph{-p}.  This means that the value of
the momentum of one particle  determines  completely  that  of  the
other i.e there is a correlation  between  their  momenta.  There is
apparently nothing quantum  mechanical  about it. However, if
remotely placed measuring devices  are  assumed  to  be uncorrelated
then  it turns out that certain  characteristics  of correlations
between systems described quantum  mechanically  can not  be
reproduced by describing them by classical  mechanics  and
probability  theory.  In fact, the simple example  given  above  of
pairs of particles created with zero total linear  momentum  was the
one used by Einstein , Podolsky and Rosen (EPR)  [1]  to raise
issues concerned with non classical features of correlations  which
are  a matter of debate even today (see [2] and the reference
therein         for         details).
\paragraph{}
In  this  article  we intend to explain the concept of superposition
and entanglement by considering the examples of one and two spin-1/2
systems(binary systems). We also intend to see the application of
superposition and entanglement in realm of biological systems. See
also the articles in reference [3,5].
\section{2. Superposition in Spin-1/2 systems}
The  dynamical  state  of  a  particle  in  classical  mechanics  is
described by the values of its dynamical observables like position
\emph{q(t)} and momentum \emph{p(t)} as a function of time. If the
particle is moving in a potential  of  known  form  then,  for given
initial conditions, the values  of the observables can be predicted
with certainty, at least in principle. However, if the force acting
on the particle is random then  it may not be possible to predict
with certainty the values of all  of its observables. A familiar
example of this situation is the motion  of  atoms constituting a
gas. Every atom in a gas is subject to  collisions  with  other
atoms. The collisions are random giving rise  to  randomness  in the
motion of  the  atoms. The values of dynamical  observables
\emph{q}, \emph{p}  of an atom  in  this  case can not be evaluated
deterministically. They are  described statistically in terms  of a
probability distribution function \emph{f(q,p)} where
\emph{f(q,p)dqdp}  gives the  probability  that  the  position  and
momentum  of  an  atom in the  gas  lie  in the interval \emph{q},
\emph{q+dq}  and \emph{p}, \emph{p+dp}. The theory predicts average
values of observables. The theoretical predictions are compared with
experiments  by  carrying measurement of the observables on a large
number  of  identical systems and by averaging the outcomes of such
measurements.
\paragraph{}
In  contrast  with  the  classical  theory,  quantum theory does not
assign  definite  values simultaneously to all the observables, like
position and momentum, even in potentials of definite form. It makes
only  statistical  predictions  about  the values of the obervables.
Hence,  quantum  theoretic  predictions  are  compared  with average
result  of  identical  experiments  performed  on identical systems.
\paragraph{}
In  quantum  theory, an isolated system is represented by the vector
denoted,   following  Dirac,  by  the  symbol  $|\rangle$  called  a
\textsl{ket.}  Thus a particular state of a system may be denoted by
$|\psi\rangle$. The vector denoting a state is also referred to as a
state  vector.  The  set  of  all  the  vectors representing all the
possible  states of a system is said to constitute the \textsl{state
space}.  The  Dirac  notation may be understood, for example, in the
familiar  language  of  vectors  by  thinking of the familiar vector
notation  \textbf{a}  or  as a ket, say $|a\rangle$. In analogy with
the familiar three-dimensional configuration space, the vector space
of  the  states  of  a  system is assigned the following properties:
\section{}
\begin{itemize}
\paragraph{}    \item Like the \textit{scalar product}\textbf{ a.b} between
two   vectors   \textbf{a   }and   \textbf{b},   we   define  scalar
product between the vectors in the state space of the system. The
scalar product between the vectors $|\psi\rangle$ and $|\phi\rangle$
is denoted by $\langle\psi|\phi\rangle$ where $\langle |$ is called
a \textit{bra}. The number $\langle\psi|\phi\rangle$ may be complex.
The scalar product is defined so that $\langle\psi|\phi\rangle$ =
$\langle\phi|\psi\rangle$
\paragraph{}    \item The number $ \sqrt{\langle\psi|\psi\rangle} $ is called the
length    or    the    norm    of    the    vector   $|\psi\rangle$.
\item  Two  vectors $ |\psi\rangle $ $ |\phi\rangle $ are said to be
orthogonal   to   each   other   if   $\langle\phi|\psi\rangle$   =0
\paragraph{}
\item   Like   any   other  vector  in  ordinary  three  dimensional
space   may   be  represented  as  a  linear  combination  of  three
basis  vectors,  a  vector  in  then  may be represented in terms of
a  complete  set  of  vectors  called  a  \emph{basis}.  A  basis is
constituted  by  any  set  of  linearly  independent  vectors in the
state  space.  The  number  of  linearly  independent  vectors  in a
given      state      is      called      its      \emph{dimension.}
\paragraph{}
\item  A  basis  may  be  constructed  by  identifying  the possible
values   that   an   arbitrarily  chosen  observable  of  the  given
system  can  take.  There  may  be  more  than  one observables that
can   be   measured   simultaneously  to  any  degree  of  accuracy.
However,  for  the  sake  of  definiteness,  we assume that only one
observable,  say  $\hat{O}$,  can  be  measured  at  a  time  to any
degree  of  accuracy.  Let $ a_{1},a_{2},....a_{m}$ be the complete
set of values  of  $ \hat{O}$  with  m  as  the  dimension  of  the
given state    space.    The   number  $ a_{i}   (i=   1,2....m)$,
being observables, are  necessarily  real.  Let  $|a_{i}\rangle$ be
the state  of  the  system  when it is known to have the value
$a_{i} (i= 1,2....m)$  for  the  observable  $\hat{O}$.  It can be
shown that the  set  of  vectors $ |a1\rangle,|a2\rangle,......,
|a3\rangle$

\begin{eqnarray}
 \langle  a_{i}|a  _{j}\rangle  = \delta _{ij}
\end{eqnarray}

and  that  it  is  complete.  Hence,  any vector in the space can be
expressed    as    a    linear    combination    or   superposition.
\begin{eqnarray}
  |\psi\rangle  = \sum_{i} C_{i}|a_{i}\rangle
\end{eqnarray}

The  set  of  vectors  $\langle a_{i}|$ (i=1,2....m) thus
constitutes a
basis.
\paragraph{}    \item On taking the scalar product of (2)with $ \langle a_{j}|$
and  using  (1), the expansion coefficients $ C_{i} $ may be seen to
be                              given                             by

\begin{eqnarray}
      C_{j}      =      \langle      a_{i}|\psi\rangle
\end{eqnarray}

\paragraph{}    \item  If a state vector is represented as in (2) then its
corresponding          bra         is         represented         as

\begin{eqnarray}
  \langle\psi|=\sum_{i=1}^{m}C_{i}^{*}\langle a_{i}|
\end{eqnarray}

\paragraph{}  \item  If (1) holds then, invoking also (4), it is
straightforward   to   see  that  the  scalar  product  between  two
vectors            may            be            written           as

\begin{eqnarray}
\langle\phi|\psi\rangle= \sum_{i=1}^{m}D_{i}^{*}C_{i}
\end{eqnarray}
where

\begin{eqnarray}
|\phi\rangle=\sum_{i=1}^{m}D_{i}|a_{i}\rangle
\end{eqnarray}
\paragraph{}    \item If $ |\psi\rangle $ representes a state and if c is a complex
number  then,  according  to  the quantum theory, c $ |\psi\rangle $
represents  the  same  state.  We  may,  therefore represent quantum
states   by   the  vectors  whose  norm  is  unity  i.e  by  vectors
$|\psi\rangle$  such  that  $\langle  \psi|\psi\rangle$  =  1. In
what follows  we  will  assume  that  the  kets  are normalized to
unity. Hence  if  a  state  $|\psi\rangle$  is represented as in
(2)then the normalization       condition       demands       the
relation
\begin{eqnarray}
\sum_{i=1}^{m}|C_{i}|^{2}=          1
\end{eqnarray}
between            the            expansion            coefficients.
The  complex  number  $  C_{i}$  in  the  expansion (2)is called the
\emph{probability   amplitude}   for  the  system  to  be  in  state
$|a_{i}\rangle$.  What  is  the physical significance of probability
amplitude?
\paragraph{}
The   physical   significance   of   the  probability  amplitude is
contained  in  the  so  called  \emph{measurement  problem}  of the
quantum  theory.  According  to  it,  if  the  possible outcomes of
measurement   of   an   observable   on   a   system are  $  a_{1},
a_{2}....a_{m}        $        with        the vectors       $
|a_{1}\rangle,|a_{2}\rangle,.....|a_{m}\rangle$ denoting   the
corresponding  states,  and  if  the  outcome  of a particular act
of measurement is, say $a_{i}$ then after the measurement is over,
the system goes over to  or  collapse  to the state $|a_{i}\rangle$.
Moreover,  if we repeat  the  same experiment  on several similarly
prepared systems then  the probability  that the result of
measurement is $a_{i}$ is given by $|C_{i}|^{2}$. In other words, if
we perform measurements of the observable $\hat{O}$ on a number M of
systems all of which are in the state given  by  (2)then  the
outcome  of  each measurement will be  a number from the set of real
numbers $a_{1}, a_{2}....a_{m}$. Let $n_{i}$  be  the number of
systems which give $a_{i}$ as the result of measurement. Then, in
the  limit $M\rightarrow\infty$,
\begin{eqnarray}
 |C_{i}|^{2} = n_{i}/M
\end{eqnarray}
The   average   value   of   the   observable   $\hat{O}$   then  is
\begin{eqnarray}
 \langle\hat{O}\rangle= \Sigma_{i=1}^{m}a_{i}|C_{i}|^{2}
\end{eqnarray}

The   procedure   outlined   above,  however,  determines  only  the
magnitude     $|C_{i}|$    of    the    complex    number
\begin{eqnarray}
    C_{i}=|C_{i}|exp  (i\theta)
\end{eqnarray}
but not its phase $\theta$. Determination of phase  requires
phase  sensitive  measurement.  However,  we do not discuss
that                issue               here.
\paragraph                                                        {}
It  is  thus  clear  that we can not determine the state of a system
if  we  have  only  one  copy  of it. The determination of the state
of  a  system  requires  a large collection of its identical copies.
From  this  we  conclude  that  we  can not make copies of a quantum
system  if  we  do  not know its state. For, if we have one copy of
a  quantum  system  in  an  unknown state and if we can make its
copy  then  we  can  make  any  number  of  its  copies  and perform
experiments  on  each  of  those copies to determine its state. That
amounts  to  only  one  copy  determining  the  state of the system!
Though  it  is  a  trivial  consequence of the measurement postulate
of   quantum   theory,  the  impossibility  of  making  a  copy  (or
cloning)a  quantum  system  in  an  unknown  state is referred to as
\emph{no-cloning                                          principle}
\paragraph{}
We   illustrate   various   notions   introduced  above  by  way  of
following                        two                       examples:
\paragraph{} \item Consider a particle whose spin measurement in any
direction  gives  $\pm\hbar/2$ as the outcome. Consider a collection
of   such   spins   oriented   in   the   direction   \textbf{e}.  A
measurement   in   the  direction  \textbf{e} on  any  spin  in that
collection  will  give  $\hbar/2$  as  the  answer.  We label such a
collection    by    the    state   $|1/2,\textbf{e}\rangle$. Since
$\pm\hbar/2$ are  the  only  two  possible  outcomes  of measurement
of any spin component  of  the  particle  in  question, its  basis
states  are $|\pm1/2,\textbf{e}\rangle$ where \textbf{e} is  an
arbitrarily  chosen  direction. These basis states are orthogonal to
each other i.e $\langle1/2,e|-1/2,e\rangle$ = 0. Consequently, the
state of the spin in any   direction   can   be represented  in
terms  of  the  states $|\pm1/2,\textbf{e}\rangle$. The  state space
of the system in this case         is, therefore, two dimensional.
\paragraph{}
A  spin  1/2  is  a  reality may be a two state system of interest
in Nuclear   Magnetic   Resonance  (NMR).
\paragraph{}
\item   Consider   an  atom  with  one  valence  electron.  Let  the
energies  of  the  levels  that  the electron can occupy be given by
$E_{1},  E_{2}....$  Any measurement of energy of the electron would
yield   one   of  those  values  of  energy  as  the  answer.  If  a
measurement  of  the  energy  of  the  electron gives $E_{i}$ as the
answer  then  we  say that the atom is in the state $|E_{i}\rangle$.
The   set   of   basis  vectors  in  this  case  is  $|E_{1}\rangle,
|E_{2}\rangle..$  which  means  that  the  dimension  of  the  state
space               is              countably              infinite.
\paragraph{}
A  particular  case  of  interest  is  an  atom  constrained to make
transitions  between  only  two  of  its  levels,  say,  the  levels
$|E_{1}\rangle$   and   $|E_{2}\rangle$   having   energies  $E_{1}$
and  $E_{2}$  $(E_{2}>E_{1})$.  That  can  be  achieved if the atom,
initially  in  a  superposition  of  the  states $|E_{i}\rangle$ and
$|E_{i}\rangle$    interacts   with   a   single   frequency   field
$\omega$ such  that  $\omega  \sim  (E_{2}-E_{1})/\hbar$  and  if
for all the other        energy        levels        $E_{i}$the
detunings $|\omega-|E_{i}-E_{1}|/\hbar|$ and
$|\omega-|E_{i}-E_{2}|/\hbar|$ are  sufficiently  large.  The atom
in  this  case is said to be a two   level   atom   driven on
resonance   by   the   field.[2].
\paragraph{}
Whatever   its   realization,   the   state  vector  representing  a
two-level     system    may    be    written,    following    (2),
as $|\psi\rangle      =      C_{0}|0\rangle      +
C_{1}|1\rangle$ where   the   states  $|0\rangle$  and  $|1\rangle$
are  orthogonal to   each   other.   The   states   ${|0\rangle,
|1\rangle  }$  may stand,    for    example,    for    the    spin
states   $|\pm1/2, \textbf{e}\rangle$     or    the    two    states
$|E_{1}\rangle, |E_{2}\rangle$   of   an   atom   involved   in  an
interaction  or the   two   states   of   any  other  two-level
system.  By  virtue of   (3),   the   expansion   coefficients  in
(10)  are  given  by

\begin{eqnarray}
 C_{j}=     \langle     j|\psi\rangle
\end{eqnarray}
 where $j =0,1$

These     coefficients     are, of    course,    subject    to the
normalization condition (7).

\section{3.        Superposition        versus        Mixture       }
How   is  the  probability  arrived  at  by  using  the  concept  of
probability  amplitude  different  from the notion of probability in
the  classical theory of probability? Consider a two level atom in a
superposition  state  (10).  As our discussion above indicates, that
state  represents  a  collection  of  identical  atoms such that the
valence  electron  in  each of the atoms in that collection occupies
the  energy level $E_{0}$ with the probability $|C_{0}|^{2}$ and the
energy level $E_{1}$ with the probability $|C_{1}|^{2}$. Does it not
mean  that  electrons  in a fraction $|C_{0}|^{2}$of that collection
occupy  level  $E_{0}$  and  that  the  electrons  in  the remaining
fraction  $|C_{1}|^{2}$  occupy level $E_{1}$? The answer is: No. We
can not separate the collection of atoms described by the state (10)
in  to  two  parts in one of which the atoms are in one state and in
the other state in the remaining part with the size of the two parts
determined by the above mentioned probabilities. In other words, the
expression  (10) does not assign "either this" "or that" state to an
electron. It states that the electron is "simultaneously" in the two
states. Only when we perform an experiment to determine which energy
state  the  electron  is,  in  that  we find the answer as $E_{0}$
or $E_{1}$  with  the number of atoms in one or the other energy
level determined  by the probabilities $|C_{0}|^{2}$ and
$|C_{1}|^{2}$. Of course,  once  an atom is experimentally found to
be in a particular level,   it   remains  in  that  level  till
disturbed.  After  the measurement  to determine their energy is
performed on all the atoms in the given collection, that collection
separates in two groups. In one  of the groups all the atoms are in
the level $E_{0}$ and in the other  all are in the level $E_{1}$. If
we mix these two groups then the probability of picking an atom from
this collection such that it has  energy  $E_{0}$ is $|C_{0}|^{2}$
and that of picking an atom of energy  $E_{1}$  is  $|C_{1}|^{2}$.
Though  this is the same set of probabilities  as  for  the
superposed state (10), the collection of spins  in question is
represented, not by the state (10), but by the mixed
state

\begin{eqnarray}
    \hat{     \rho}=    |C_{0}|^{2}|E_{0}\rangle\langle
E_{0}|+ |C_{1}|^{2}|E_{1}\rangle\langle             E_{1}|
\end{eqnarray}
The  operator  $\hat{\rho}$ in  the  equation  above  is called
the density  matrix.  However,  here  we  do  not discuss the
concept of density  matrix.  Each atom in the collection described
by (12) is in "either           one"           "or          other"
state.
\paragraph{}
If  you  feel  uncomfortable with the counterintuitive picture of an
electron  being  simultaneously  in  two levels (instead of being in
either  one  or the other level) till an experiment throws it in one
or the other level then you are not alone. You are in the company of
none  other  than  Einstein.  He questioned this apparently peculiar
situation  by asking: "...does it mean moon is not there till I look
at  it"?  and  "God  does not play dice". Howsoever strange it might
appear, that is how the nature is in the eyes of the quantum theory!
\paragraph{}
One might wonder, after one finds the atom in one or the other level
after  performing  an  experiment,  how  does  one  know that it was
"simultaneously"  in  the  two  levels  and  not in "either one" "or
other"  level  till  then?  The  answer  is:  we can not distinguish
between  "simultaneous" and "either"/"or" situation by an experiment
that  is  designed to determine only the level in which the atom is!
That   is  because  such  an  experiment  determines  the  magnitude
$|C_{0}|$  and  $|C_{1}|$ but  not  the  phase  of $C_{0}$  or that
of $C_{1}$. The said two situations can be distinguished by
experiments which        are        sensitive        to        the
phase.
\section{4. Classical Statistics and Quantum Mechanics of a Spin-1/2 system}
In  this section we examine the similarities and differences between
the probabilistic predictions of quantum mechanics of a spin 1/2 and
compare  them  with  the  predictions  arrived  at  by its classical
statistical  description  assuming  that  the  spin  is subject to a
random    force.    In    the    following    we   take   $\hbar=1$.
\paragraph{}
Consider a collection of identically prepared spin-1/2 each oriented
in  the  spherical polar direction $(\theta,\phi)$. Let $|+z\rangle$
denote  the  state  of a collection of spins all in the direction +z
and  let  $|-z\rangle$ denote the state of a collection of spins all
in the direction -z. Choosing these state as the basis, we may write
[2]

\begin{eqnarray}
|+;\theta,\phi\rangle= \cos(\theta/2)|+z\rangle+
\sin(\theta/2)\exp(i\phi)|-z\rangle
\end{eqnarray}

On  recalling  (2)  and  the  discussion following it, we see that a
measurement  of  the  z-component  of  spins in the state (13) would
give 1/2 as the outcome with probability $cos^{2}(\theta/2)$ and
-1/2 as  the  outcome  with  probability $sin^{2}(\theta/2)$. The
average value    of    spin   in   the   z-direction   will
therefore   be $(cos^{2}(\theta/2)-sin^{2}(\theta/2)= cos(\theta)/2$
\paragraph{}
Now,  let us visualize the spin-1/2 as a classical two-state object.
In  order to mimick the quantum results, we may assume that the spin
is  undergoing  a  random motion as a result of which its projection
along  any  direction is not fixed but a random number which assumes
the  values  $\pm1/2$ with the same probability as the corresponding
quantum  spin.  Thus,  if  we  assume  that  the quantum spin in the
direction  $(\theta,\phi)$is  represented  by a classical spin whose
projection  along  the  direction  $(\theta,\phi)$  gives 1/2 as the
outcome of measurement with probability $cos^{2}(\theta/2)$ and -1/2
as  the  outcome of measurement with probability $sin^{2}(\theta/2)$
then  the  classical  and  the  quantum  pictures emerge in complete
agreement.
\paragraph{}
There  is,  however, a catch in the argument above. For, recall from
the  Sec.2  that  statistical  aspects  are brought in the classical
picture  if the system under consideration is under the influence of
a  random  force  due  to  its environment. However, the spin in our
example   here   is   isolated.  It  is  not  interacting  with  any
environment.  How  do  we  explain  its  assumed  random behavior? A
possible  way  out  is  to  assume  that  the  spin interacts with a
fictitious  environment consisting of unknown entities called hidden
variables.  This  picture  is drawn in analogy with the motion of an
atom  in  a  gas  undergoing  random collisions with other atoms. In
that  case,  the  other  atoms  constitute  hidden  entities  for an
observer    monitoring    the   motion   of   a   particular   atom.
\paragraph{}
Thus,  if  we  assume  the  existence of a fictitious environment of
hidden  variables  then  the quantum mechanical prediction about the
average  value  of a component of a spin-1/2 in any direction can be
mimicked  by  treating  the  spin as a classical object. The average
value  of  outcome  of  measurements  of the component of a spin-1/2
in  any  direction  thus  does  not  provide  any signature that can
distinguish        quantum       and       classical       theories.
\paragraph{}
If  the  hidden  variables  theory succeeds in reproducing all those
predictions  of  the  quantum  theory  which  are  in agreement with
observations  then  the  problem  of  solving the mystery of quantum
mechanics  would  reduce to solving the mystery of hidden variables.
However,  the  system  of  two  spin-1/2s,  discussed in Sec.6 and 7
below  brings  out  the shortcomings of the hidden variable theory.

\paragraph{}
The  failure  of the hidden variable theory, assuming the process of
measurement  to  be  local,  is  attributable  to  the  property  of
entanglement                     discussed                     next.

\section{5.                       Entanglement                      }
We  have  seen that an isolated quantum system is described by a set
of  probability  amplitudes  each  for  an  admissible  value  of an
observable   or   a   set  of  observables  which  can  be  measured
simultaneously  to  any  degree  of  accuracy. Consider now a system
consisting  of  two  subsystems  denoted  by  A and B. In principle,
quantum  theory  permits  measurement of any observable of A and any
observable  of  B  simultaneously  to  any  degree  of accuracy. Let
$a_{1},a_{2},.....a_{m}$ be the complete set of values of one of the
observables                 of                 A                with
$|a_{1}\rangle,|a_{2}\rangle,....|a_{m}\rangle$     denoting     the
corresponding states. Similarly, let $b_{1},b_{2},.....b_{m}$ be the
complete  set  of  values  of  one  of  the  observables  of  B with
$|b_{1}\rangle,|b_{2}\rangle,....|b_{m}\rangle$     denoting     the
corresponding  states. Any state of the combined system is evidently
described  by  the probability amplitudes $C_{ij}$ for observing the
set  of  mn  amplitudes $(a_{i},a_{j})  (i=1,2,...m;j=1,2,...n)$.
The number $|C_{ij}|^{2}$ is the probability for A to have value
$a_{i}$ when  the value of B is $b_{j} (i=1,2...m;j=1,2,...n)$ as
the outcome of  the  measurement of corresponding observables. The
corresponding combined    states   are   denoted   by
$|a_{i},b_{j}\rangle\equiv |a_{i}\rangle|b_{j}\rangle$. A general
combined state $|\psi\rangle$ is represented
by

\begin{eqnarray}
  |\psi\rangle  =  \Sigma_{i,j} C_{ij} |a_{i},b_{j}\rangle
\end{eqnarray}

Now,  let  the  probability  amplitudes  $C_{ij}$ be such that
they reduce        (14)        to        the factorized form
\begin{eqnarray}
     |\psi\rangle    =    |\psi_{a}\rangle |\psi_{b}\rangle
\end{eqnarray}

where   $|\psi_{a}\rangle$   is  a  state  of  system  A  alone
and $|\psi_{b}\rangle$  is the that of the system B alone. The
form (15) implies  that  the  systems  A  and B are uncoupled or
uncorrelated.

\paragraph{}
However, if the combined state of the two systems A and B can not be
factorized (as in (15)) in to a product having a state only of A and
a  state  only  of  B  as its factors then it is called an entangled
state.  We  discuss the significance of entangled states by means of
the   example   of   two   spin-1/2s   in  the  following  sections.

\section{6.                       Two                      Spin-1/2}
Consider  a  system  consisting of two spin 1/2s denoted by A and B.
Following  Sec.5  we  can  specify a state of the system in terms of
values of the component of spin A along some direction $\vec{e_{A}}$
and   those  of  the  components  of  spin  B  along  the  direction
$\vec{e_{B}}$where  $\vec{e_{A}}$  and  $\vec{e_{B}}$ may be same or
different  directions. The possible states of such a system then are
linear     combinations     of     the     states    $|+\vec{e_{A}},
+\vec{e_{A}}\rangle,              |+\vec{e_{A}},-\vec{e_{B}}\rangle,
|-\vec{e_{A}},+\vec{e_{B}}\rangle,
|-\vec{e_{A}},-\vec{e_{B}}\rangle$                             where
$|\epsilon_{a}\vec{e_{A}},\epsilon_{b}\vec{e_{B}}\rangle      \equiv
|\epsilon_{a}\vec{e_{A}}\rangle|\epsilon_{b}\vec{e_{B}}\rangle$
(where$\epsilon_{a},  \epsilon_{b} = \pm1$) denotes the state of the
combined  system  in  which  the  components  of  spin  A  along the
direction $\vec{e_{A}}$ is $\epsilon_{a}/2$ and that of spin B along
the  direction  $\vec{e_{B}}$  is $\epsilon_{B}/2$. Any state of the
combined  system  of  two  spin  1/2s  may therefore be expressed as

\begin{eqnarray}
 |\psi\rangle    =   \alpha_{1}|+\vec{e_{A}},
+\vec{e_{A}}\rangle+ \alpha_{2} |+\vec{e_{A}},-\vec{e_{B}}\rangle
+\nonumber\\ \alpha_{3}|-\vec{e_{A}},+\vec{e_{B}}\rangle +
\alpha_{4}|-\vec{e_{A}},-\vec{e_{B}}\rangle
\end{eqnarray}
Clearly,  $|\alpha_{1}|^{2}$gives the probability that the result of
simultaneous measurement of the component of spin A in the direction
$\vec{e_{A}}$  is  1/2  and  that  of the component of spin B in the
direction  $\vec{e_{B}}$ is also 1/2. Similar interpretation applies
to other terms in (16). The state (16) will be an entangled state if
we  can  not  write  it  in  the  form (15) with $|\psi_{A}\rangle$
representing  a  state  of A alone. As an example of a non-entangled
state,   note   that   if $\alpha_{i}=   0$  for  $i\neq  2$  then
$|\psi\rangle$  in (16) is of the form (15) with $|\psi_{A}\rangle =
|+\vec{e_{A}}\rangle$ and $|\psi_{B}\rangle = |+\vec{e_{B}}\rangle$.
As   another example,   if   $\alpha_{i}=  1/2$  for  all  i  then
$|\psi\rangle$  in (16) can be expressed in the factorized form (15)
with      $|\psi_{A}\rangle      =      (|+\vec{e_{A}}\rangle +
|-\vec{e_{A}}\rangle)/\sqrt{2}$      and     $ |\psi_{B}\rangle =
(|+\vec{e_{B}}\rangle  +  |-\vec{e_{B}}\rangle)/\sqrt{2}$. Consider
now the state

\begin{eqnarray}
|\psi\rangle = 1/\sqrt{2} [|\vec{e},-\vec{e}\rangle -
|-\vec{e},\vec{e}\rangle]
\end{eqnarray}

It  may  be  verified  that  this  state can not be factorized as in
(15).  Hence  it  is  an  entangled  state. Since the coefficient of
$|\vec{e},-\vec{e}\rangle$ and that of $|-\vec{e},\vec{e}\rangle$ in
(17)  is zero, it follows that (17) represents a state for which the
probability   of   finding   both  the  spins  aligned  parallel  or
antiparallel          to          $\vec{e}$         is         zero.

\paragraph{}
The  question  that  may  be  asked is: Can we measure the extent of
entanglement?   In   other  words,  can  we  formulate  a  criterion
according  to which we can compare the extent of entanglement of two
states  and  say that one state is more entangled than the other? It
turns   out   that   it  is  possible  to  construct  a  measure  of
entanglement  of  a  state of a system to two spin-1/2s described in
terms  of  a  state vector as in (16). The state (17) is accordingly
found   to   be   maximally   entangled   (see   [2]  for  details).
\section{7.          Signature          of          Non-Classicality}
In  order  to  understand  the  meaning  of  non-classical features,
consider the two spin-1/2s prepared in the maximally entangled state
(17). Let those spins fly apart. After they are separated and do not
interact  any  longer,  measure  the component  of  spin  A in some
direction  $\vec{a}$  and  that  of spin  B  in  another  direction
$\vec{b}$  ($\vec{a}$  and $ \vec{b}$  are  unit  vectors). Repeat
the experiment   N   times. These   measurements   give   the
numbers $N(+\vec{a};+\vec{b}),
N(+\vec{a};-\vec{b}),N(-\vec{a};+\vec{b}),N(-\vec{a};-\vec{b})$
where $N(+\vec{a};+\vec{b})$ is the number of times the result of
the said  joint  measurement  on  A  is  1/2 when that on B is also
1/2; $N(+\vec{a};-\vec{b})$ is  the number of times the result of
the said joint  measurement  of  A  is  1/2 when that on B is -1/2
and so on. These       measurements       determine      the
probabilities $p(\epsilon_{a}\vec{a};\epsilon_{b}\vec{b})=
N(\epsilon_{a}a;\epsilon_{b}b)/N$ where
$p(\epsilon_{a}\vec{a};\epsilon_{b}\vec{b})$ is  the probability
that the  outcome  of joint measurement of the component of spin A
in the direction  $\vec{a}$  is $\epsilon_{a}/2$ when that of the
component of  spin  B  in the direction $\vec{b}$ is
$\varepsilon_{b}/2$ (with $\epsilon_{a}= \pm1,  \epsilon_{b}=
\pm1$). If the spins are in the state   (17) then   the
quantum-theoretic  expression  for  this probability is     found to
be     given     by     [2]

\begin{eqnarray}
p(\epsilon_{a}a;\epsilon_{b}b)=        1/4        [       1 -
\epsilon_{a}\epsilon_{b}\vec{a}.\vec{b}]
\end{eqnarray}

From  this we infer that the probability of finding two spins in the
same   direction   is   zero,   i.e  if  $\vec{a}.\vec{b}=  1$  then
$p(+\vec{a};+\vec{b})=  p(-\vec{a};-\vec{b})=  0$. This implies that
if  a spin is found to be aligned in any direction $\vec{a}$ then we
know  that  the  other  spin  is  aligned  along $-\vec{a}$. This is
evidently   consistent   with   the   discussion   following   (17).

\paragraph{}
From  the point of view of the discussion to follow, we consider the
probabilities  for  the  pairs  of  directions  from  a set of three
directions  $\vec{a},\vec{b}$  and  $\vec{c}$.  Use  (18) to show
that

\begin{eqnarray}
p(+\vec{a};+\vec{b})+ p(+\vec{b};+\vec{c})-p(+\vec{a};+\vec{c})=
\nonumber\\1/2 [sin^{2}(\theta_{ab}/2)+  sin^{2}(\theta_{bc}/2)-
sin^{2}(\theta_{ac}/2)]
\end{eqnarray}

where  $\theta_{ab}, \theta_{bc}, \theta_{ac}$are the angles between
the  directions identified by the respective subscripts. The Eq.(19)
is             quantum             theoretic             prediction.

\paragraph{}
Let  us  now examine the probabilities on the left hand side of (19)
by  treating  the  two spins as classical two-state objects. In this
picture,  each  of the spin component can assume the values $\pm1/2$
with yet to be specified probabilities. An important identity in the
classical  description  involving the probabilities appearing on the
left hand side of (19) is found by expressing those probabilities in
terms  of joint probabilities in three directions as explained next.
To that end, consider three directions $\vec{a},\vec{b},\vec{c}$ and
define a joint probability for the components of the two spins along
those directions to have specific values. For example, we define the
joint                                                    probability
$p(\vec{a},\vec{b},\vec{c};\vec{a},\vec{b},\vec{c})$     for     the
components  of both the spins along the given three directions to be
1/2  (the quantities to the left of the semicolon in p refer to spin
A and those to its right refer to spin B). We can in general, define
the                        joint                       probabilities
$p(\pm\vec{a},\pm\vec{b},\pm\vec{c};\pm\vec{a},\pm\vec{b},\pm\vec{c})$.

\paragraph{}
Note   that   the  joint  probabilities  give  the  probability  for
components  in  different  directions  of the spins to have definite
values  simultaneously.  We can, of course, measure any component of
one   spin   and   the  same  or  any  other  component  of  another
simultaneously  to  any  degree  of  accuracy.  However,  in quantum
formalism,  we  can  not  assign  definite  values simultaneously to
different  components of same spin. Hence, there is no place for the
joint  probabilities for definite values of the components of a spin
in  different  directions in quantum formalism. Hence, following the
notation  introduced  above  whereby the entries on two sides of the
semicolon  in p refer to different spin,
$p(\pm\vec{a};\pm\vec{b})$is a legitimate quantum mechanical
probability but any p in which there are  more  than  one  entries
on  any  side  of  the  semicolon  is inadmissible in quantum
theory. The consequences arrived at below by invoking  the  notion
of joint probability are, therefore, strictly classical.

\paragraph{}
Next,  recall  that the probability $p(x_{1},x_{2},....x_{m})$ for m
variables  $x_{1},x_{2},....x_{m}$  may  be  obtained by summing (or
integrating)over  all  the  permissible  values  of  the n variables
$x_{m+1},x_{m+2}.....x_{m+n}$        in        the       probability
$p(x_{1},x_{2},....x_{m},x_{m+1},x_{m+2}.....x_{m+n})$     of    m+n
variables.  Hence,  the probabilities for two variables appearing on
the  left  hand  side of (19) may be expressed in terms of the joint
probabilities  for  three  directions introduced above. For example,
\begin{eqnarray}
p(+\vec{a};+\vec{b})=\sum
p(+\vec{a},\epsilon_{1}\vec{b},\epsilon_{2}\vec{c};
\epsilon_{3}\vec{a},+\vec{b},\epsilon_{4}\vec{c})
\end{eqnarray}

where  the summation is over the $\epsilon$'s each taking the
values $\pm1$.

\paragraph{}
Now,  we  invoke  (1)  the  key property of the entangled state (17)
embodied  in  (18)  namely that the probability of finding two spins
aligned parallel to each other is zero, (2)the relations to the type
(20)   for   the  probabilities  appearing  in  (19),  and  (3)  the
requirement  that  all  the  probabilities  be  non  negative. These
conditions       lead       to       the       inequality      [2,4]

\begin{eqnarray}
p(+\vec{a};+\vec{b}) +
p(+\vec{b};+\vec{c})-p(+\vec{a};+\vec{c})\geq   0
\end{eqnarray}

called a \emph{Bell's inequality}. Recall that the corresponding
quantum theoretic  result  is  the  equality  (19).  That
equality need not respect  the  inequality  (21).  A violation of
(21) by (19)for some choice  of  directions  would constitute a
rebuttal of the classical description.
\paragraph{}
It  is  not difficult to see that if, for example, the three vectors
are        coplanar        and       $\theta_{ab}=\theta_{bc}=\pi/3,
\theta_{ac}=2\pi/3$,  then  the value of (19) is -1/4 which violates
(21).  Several  other  inequalities  have been similarly derived for
other situations and the experiments carried to confirm violation of
Bell's                                                 inequalities.
\paragraph{}
Note     from     the     discussion     preceding     (21) that of
the three assumptions leading to the Bell's inequality, the first
one  concerns  the  preparation of the system in a particular state.
The  fact that such a state is a practical reality, the violation of
(21)  must  be  due  to  the  failure of  one  or both of other two
assumptions.  The  second assumption is based on the possibility of
defining  joint probabilities  for  different  components  to  have
definite values.  We  may,  therefore,  attribute  the violation of
(21)as confirming  that  the  joint  probabilities  for  different
components of a spin to have definite values is indeed inadmissible.
However,   if   we   insist   on  retaining  the concept  of  joint
probabilities  of  all kinds then the only reason for the failure of
(21)  must be the failure of the assumption (3)which states that the
probabilities  must  be positive.  Its  failure  implies  that  the
probabilities  are not necessarily positive. The concept of negative
probabilities, however, has no place in the classical theory. Hence,
violation of  a  Bell's  inequality  in  any case is a signature of
non-classicality.

\paragraph{}An  entangled  state  thus  exhibits  features  having  no classical
counterpart.  Note  that if the state is not entangled i.e if it can
be factorized as in (15) then each of the spin behaves independently
of  the other. We have seen in Sec.4 that a single spin-1/2 does not
exhibit  any  non-classical property as regards the averages of spin
components.  Hence  two  independent spins  can not exhibit any non
classical  property  implying thereby that a factorizable spin state
may  be  mimicked classically.  Recall also that though the quantum
averages of a single spin in a superposition state can be reproduced
by  its classical statistical description, that description can not
reproduce the properties resulting from interference between the two
states.\\The  pillars  of QIP, namely superposition and entanglement
thus can not      be     built     by     any classical
prescription!
\section{8. Application to Quantum Life units}
 If we want to go into a deep understanding of the
phenomenon of superposition and entanglement, the question that
comes forth is: Whether this phenomenon of super position and
entanglement is restricted to quantum domain or it plays an
important role in the dynamics of living units. Inanimate objects in
macroscopic world are totally deterministic, the probability arising
in the macroscopic world is only due to the lack of knowledge of the
system. If we consider a simple example of coin tossing the chance
factor that comes into play because we are totally ignorant of the
parameters like air resistance, bias ness of the coin, etc. On the
contrary quantum mechanics, representing the microscopic system, is
totally probabilistic, and this probability is not because of the
lack of the knowledge of the system but fundamentally due to the
uncertainty in the system given by Heisenberg's Uncertianty
principle.\\If we analyze into the dynamics of living units, what is
then the underlying physics of living units, 'Is it classical
mechanics or quantum mechanics?' Even if we try to find out a
correspondence between the dynamics of the living system with that
of the non living systems, still there remains some extensive
biological processes which are difficult to correlate with the
quantum or classical world. Though the processes like replication,
culling of living species find different interpretation in quantum
world, but still there are many processes which are not explained in
terms of quantum theory. Similarly it is important to ask if
physical phenomenon like superposition and
entanglement have certain meaning in biological world or not.\\
Even if we assume that the phenomenon of superposition is
permissible in the biological domain, then how are we going to
explain the phenomenon of cloning which is allowed for living units?
This can be probably understood as follows: In the quantum world the
initial state of the system is $|\psi(0)\rangle=|\xi\rangle$ then it
undergoes evolution in a linear super position of a basis states
such as $\sum_{\xi} \psi_{\xi}(l)|\xi\rangle$ and after a certain
time period $t=T$ it comes back to the original state. Then the
quantum states representing artificial living systems can be copied
at certain time periods like $t=0,T,2T,.....$ and so on. If the
living organisms are microscopically small and if quantum states are
represented as artificial living systems, superposition do prevail
at this level of the living organisms
[5].\\
Regarding another important non classical feature of 'Entanglement',
this can be related with the property of interdependency between
different species which is shown in [5], process of entangling the
several copies of the living organisms with one of the mutated
version.

\section{9. Conclusion}
 Thus pillars of QIP namely superposition and
entanglement play an important role in our understanding of physics
at the quantum scale. Its application in areas like information
processing and biological systems carries a lot of promise for
technological progress and conceptual understanding. The pedagogical
understanding of QIP primitives by physical spin-1/2 systems gives a
new outlook to understand how quantum information derives its
meaning from physics of the quantum. Also it motivates us to look
for more physical systems as resources in information processing.

\end{itemize}

\textbf{Acknowledgements} Authors acknowledges Prof. R R Puri for
his stimulating way of describing Quantum Information Processing
primitives. I. Chakrabarty acknowledges Prof B.S. Chowdhuri, Prof
C.G. Chakraborti, Prof A.K.Pati for their all time support in
carrying out research.

\section{References}
[1]  A.Einstein,  B.Podolsky  and N.Rosen, Phys. Rev. 47, 777 (1935)

[2]  Ravinder  R.  Puri,  Mathematical  Methods  of  Quantum  Optics
(Springer-Verlag,         Heidelberg,         Germany,         2001)

[3]  R.R  Puri,Mystery  of quantum entanglement, Physics News, 32 38
(Jan-June  2001);  Quantum  Computing,  ibid, 33 26 (Jan-March 2002)

[4]  E.P.  Wigner,  Interpretation  of quantum mechanics, in Quantum
Theory  of  Measurement  edited  by  J  A  Wheeler  and  W  H Zurek,
(Princeton       University       Press,      Princeton,      1983).

[5] A.K.Pati, Replication and Evolution of Quantum
Species,quant-ph/ 0411075



\end{document}